\documentclass[fleqn,usenatbib]{mnras}

\usepackage{newtxtext,newtxmath}
\usepackage[T1]{fontenc}
\DeclareRobustCommand{\VAN}[3]{#2}
\let\VANthebibliography\thebibliography
\def\thebibliography{\DeclareRobustCommand{\VAN}[3]{##3}\VANthebibliography}
\usepackage{ae,aecompl}
\usepackage{natbib} 
\usepackage{amsmath}
\usepackage{graphicx}

\title[GCs contribute to the NSC and galaxy center $\gamma$-ray excess]{Globular Clusters Contribute to the Nuclear Star Cluster and Galaxy Center $\gamma$-Ray Excess, Moderated by Galaxy Assembly History}

\author[Y. Gao. et al.]{
Yuan Gao$^{1,5}$,
Hui Li$^{2,3}$\thanks{E-mail: hliastro@tsinghua.edu.cn},
Xiaojia Zhang$^{4,5}$,
Meng Su$^{1,5}$,
Stephen Chi Yung Ng$^{1,5}$
\\
$^{1}$Department of Physics, The University of Hong Kong, Pokfulam Road, Hong Kong SAR, China\\
$^{2}$Department of Astronomy, Tsinghua University, Beijing 100084, China\\
$^{3}$Department of Astronomy, Columbia University, Mail Code 5246, 538 West 120th Street, New York, USA\\
$^{4}$Department of Earth Sciences, The University of Hong Kong, Pokfulam Road, Hong Kong SAR, China\\
$^{5}$Laboratory for
Space Research, The University of Hong Kong, Cyberport 4, Hong Kong SAR, China\\
}

\date{Accepted 15-Nov-2023. Received 13-Nov-2023; in original form 05-Sep-2023}

\pubyear{2023}

\begin{document}

\label{firstpage}
\pagerange{\pageref{firstpage}--\pageref{lastpage}}
\maketitle

\begin{abstract}
Two unresolved questions at galaxy centers, namely the formation of the nuclear star cluster (NSC) and the origin of the $\gamma$-ray excess in the Milky Way (MW) and Andromeda (M31), are both related to the formation and evolution of globular clusters (GCs). They migrate towards the galaxy center due to dynamical friction, and get tidally disrupted to release the stellar mass content including millisecond pulsars (MSPs), which contribute to the NSC and $\gamma$-ray excess. In this study, we propose a semi-analytical model of GC formation and evolution that utilizes the Illustris cosmological simulation to accurately capture the formation epochs of GCs and simulate their subsequent evolution. Our analysis confirms that our GC properties at $z=0$ are consistent with observations, and our model naturally explains the formation of a massive NSC in a galaxy similar to the MW and M31. We also find a remarkable similarity in our model prediction with the $\gamma$-ray excess signal in the MW. However, our predictions fall short by approximately an order of magnitude in M31, indicating distinct origins for the two $\gamma$-ray excesses. Meanwhile, we utilize the catalog of Illustris halos to investigate the influence of galaxy assembly history. We find that the earlier a galaxy is assembled, the heavier and spatially more concentrated its GC system behaves at $z=0$. This results in a larger NSC mass and brighter $\gamma$-ray emission from deposited MSPs.  
\end{abstract}


\section{INTRODUCTION}
\label{sec intro}
A compact bright star cluster is commonly observed at the centers of galaxies of all types, known as the nuclear star cluster (NSC) (e.g. \citet{1974ApJ...194..257L,1985ApJ...292L...9K, 1997AJ....114.1899M, 2005AJ....130...73H}). They take up the innermost a few to tens of parsecs at most, and have masses $10^5 \sim 10^8\,M_{\odot}$ \citep{2016MNRAS.457.2122G,2018ApJ...869...85S}, making them the densest known star clusters. Observationally, they are distinguished by prominently brighter luminosity on top of the disk or bulge component (e.g. \citet{2002AJ....123.1389B, 2004ChJAA...4..299K}). Besides, some NSCs are also observed to co-exist with a supermassive black hole (SMBH) at the galaxy center (e.g. \citet{2003ApJ...588L..13F,2019ApJ...872..104N}). 

NSCs consist of a mixed stellar population in terms of age, metallicity, etc (e.g. \citet{2001AJ....121.1473B, 2018MNRAS.480.1973K}). The complexity in stellar population has complicated the quest of NSC formation mechanisms. The young and metal rich stars suggests an in-situ formation scenario, where local star formation is triggered by the inflow of gas induced by various mechanisms, such as bar-driven infall \citep{1990Natur.345..679S} and the action of instabilities \citep{2004ApJ...605L..13M}. The fact that this young stellar population is usually flattened, rotating, and concentrated at the center of NSCs \citep{2014MNRAS.441.3570G, 2015AJ....149..170C} also favors the in-situ formation scenario. On the other hand, the old and metal poor population can naturally arise from massive globular clusters (GCs) which migrate into the galaxy center due to dynamical friction. This was actually the first proposed NSC formation scenario \citep{1975ApJ...196..407T} one year after the ground-breaking observation of the NSC in M31 \citep{1974ApJ...194..257L}. Observationally, this scenario is also supported by evidences such as the deficit of massive GCs \citep{2009A&A...507..183C} and the nucleation fraction tracing the fraction of galaxies that have GCs \citep{2019ApJ...878...18S}. 

Despite observational motivation for both NSC formation mechanisms, a comprehensive modeling is lacking. For in-situ formation, direct simulation is challenging. There are limited studies focusing on different aspects of the process, such as gas inflow  \citep{2010MNRAS.405L..41H}, momentum feedback and self-regulation \citep{2006ApJ...650L..37M}, and stellar two-body relaxation \citep{2015ApJ...799..185A}. Ex-situ formation was more extensively studied (e.g. \citet{1975ApJ...196..407T, 1993ApJ...415..616C, 2001ApJ...552..572L, G+14, 2022MNRAS.514.5751L}). In particular, \citet{2011MNRAS.418.2697H, 2012ApJ...750..111A, 2017MNRAS.464.3720T} used direct n-body simulation to study the final parsec-scale evolution of spiraled-in GCs and resulting NSC morphological and kinematic properties. The general picture has been well-established, but previous studies adopted crude treatments of the initial conditions of GCs due to a lack of knowledge on their formation. Besides, the evolution of the old GC systems is closely correlated with the evolution of the host galaxy, which hasn't been taken good care of in previous studies. Up to today, it is still uncertain the roles of in-situ and ex-situ formation mechanisms (e.g. \citet{ 2016MNRAS.461.3620G,2022A&A...658A.172F}), with ongoing debates on topics such as a potential transition between the two mechanisms indicated by the NSC mass or galaxy stellar masses \citep{2019A&A...629A..44L}. 

In this study, we focus on the ex-situ formation of NSCs, as the infall of GCs might contribute to another unsolved problem. A diffuse $\gamma$-ray excess has been observed at the centers of the Milky Way (MW) and Andromeda (M31) galaxies \citep{A+14, 2017ApJ...836..208A}. These excesses exhibit spherical symmetry and extend over a few parsecs, with a peak energy around a few GeV. Possible origins of these excesses are debated primarily over DM (e.g. \citet{PhysRevD.91.063003, 2018PhRvD..97j3021M,2019PhRvD..99l3027D,2022PhRvD.105j3023C}) and millisecond pulsars (MSPs) (e.g. \citet{2015ApJ...812...15B,2017JCAP...05..056H,2018ApJ...862...79E,2019RAA....19...46F,F+19M31,2022MNRAS.tmp.2330Z}. 

While the spatial distribution of DM is already expected to peak at galaxy centers (e.g. \cite{NFW}), MSPs are not commonly observed there due to difficulties in resolving individual $\gamma$-ray sources. However, MSPs are abundant in GCs, with much more observed number per unit mass compared to the galaxy field \citep{2008IAUS..246..291R,2015ApJ...812...15B, 2019ApJ...877..122Y}. This is due to the fact that MSPs are believed to originate from low-mass X-ray binaries (LMXBs), where the neutron star gets spun up through mass transfer from the companion star \citep{1991PhR...203....1B}. Thus, the high stellar density of GCs provide the desirable environment for both primordial binary formation and dynamical encounter. Galaxy center MSPs can originate from GCs that have migrated in and tidally dissolved.  

On the other hand, the central bulge region of galaxies is also a dense stellar environment, although less dense by an order of magnitude than most GCs \citep{2007MNRAS.380.1685V}. Thus, MSPs could in principle form in-situ as well. However, recent studies using scaling relations to probe in-situ MSP luminosity cast doubts on this mechanism as the sole origin of the $\gamma$-ray excess. The Galactic center excess (GCE) has been examined by \citet{2015JCAP...06..043C} and \citet{2017JCAP...05..056H}, who pointed out that LMXBs are too rarely observed in the MW bulge that in-situ MSPs can account for less than a quarter of the excess luminosity. For M31, where $\sim 10^4$ LMXBs are needed to explain the excess, less than 80 were detected within the inner 12 arcmin ($\sim$ 2.7 kpc) \citep{2007MNRAS.380.1685V}. Consequently, it is crucial to explore the contribution of ex-situ MSPs from GCs.  

From the previous analysis, we see that the evolution of GCs inevitably contributes to NSC formation and the $\gamma$-ray excess, and it is important to find out the extent of its contributions. However, the formation of GCs still remains highly uncertain. Many previous studies (e.g. \citet{G+14,F+19M31,2022MNRAS.514.5751L}) assumed that all GCs formed at a single redshift, which only serves as a primitive approximation. To achieve a more reliable modeling of GC evolution and mass deposition, a better formulation of GC formation is needed.

In this paper, we use a new semi-analytic model of GC formation and evolution to study its contribution to the NSC formation and galaxy center $\gamma$-ray excess in galaxies similar to the MW and M31. The model adopts the GC formation scenario by \citet{LG14, CGL18,CG19}, where GC formation was triggered by periods of rapid mass accretion onto the host galaxy across its assembly history, typically triggered by major galactic mergers. To obtain realistic galaxy merging histories, results from the Illustris cosmological simulation are used, and GCs are sampled at qualified simulation snapshots. After formation, the GC population is subject to orbital migration and mass loss depending on their mass and galactocentric distance. In addition, we also model an evolving background potential according to the galaxy assembly history. Through this new model, we hope to enhance our understanding of the connection between GCs,the NSC and galaxy center $\gamma$-ray excess .

We arrange this paper as follows. We introduce our modeling of GC formation and evolution in Section \ref{sec methods}. In it we also show how to account for the galaxy center MSP luminosity at $z=0$. The calculation of halo parameters from Illustris outputs is introduced in the Appendix \ref{sec apdx}. In Section \ref{sec results}, we present our model predictions of GC properties, the NSC mass and $\gamma$-ray emission by MSPs at $z=0$. As we can retrieve from Illustris a collection of halos of similar masses but with different assembly histories, we discuss the moderation effect of assembly history in Section \ref{sec discus}. As our $\gamma$-ray luminosity prediction for the M31 falls short to observation, we also discuss alternative explanations. Caveats of our study are listed and discussed as well. Finally in Section \ref{sec concl}, we summarize important results and suggest future work.

\section{Methods}
\label{sec methods}
In this section, we present our semi-analytical model for the formation and evolution of GCs in the framework of hierarchical structure formation. Then we describe the calculation of the $\gamma$-ray luminosity at $z=0$ from deposited MSPs. 

\subsection{Formation of GCs in cosmological simulations}
\label{sec gcform}
We introduce our modeling of GC formation in terms of formation times, initial masses and spatial distribution. 

In modeling GC formation times, we improve on the simple prescription by previous studies (e.g. \citet{G+14,F+19M31,2022MNRAS.514.5751L}) that GCs formed at a single redshift. While this assumption is partly justified due to the old ages of most GCs, it is recently recognized that GC formation covers a wide range of cosmic time with diverse formation histories (\citet{2018RSPSA.47470616F} and references therein). Therefore, a more physically-motivated GC formation model is desired. Fortunately, over the past decades, our understanding of the origin of GCs in the framework of hierarchical structure formation has been revolutionized. Here, we adopt the GC formation model \citet{CGL18} (hereafter CGL), which was built upon earlier works by \citet{2010ApJ...718.1266M} and \citet{LG14}. The CGL model assumes that GC formation was triggered by periods of rapid mass accretion onto the host galaxy, typically during major mergers. This idea was motivated by multiple reasons such as more observed young massive clusters in interacting galaxies (e.g. \citet{ 2006ApJ...641..763W, 2010ARA&A..48..431P}), earlier formation times of GCs than the field stars and that galactic mergers were more frequent at high redshifts \citep{LG14}, and that galactic mergers are able to induce the high densities and pressures desired for cluster formation (e.g. \citet{ 2017ApJ...834...69L, 2019MNRAS.482.4528E}). 

In the CGL model, GC formation is painted onto the halo merger trees of the Illustris simulation, which captures the evolution of halo properties from $z$=47 to 0 \citep{2014Natur.509..177V,2015A&C....13...12N}. GC formation is triggered when the specific halo mass accretion rate exceeds a threshold value, which is a tunable parameter. The total GC mass is mapped from the halo mass through observed stellar mass-halo mass relation (SMHM) \citep{2013ApJ...770...57B}, the stellar mass-gas mass relation \citep{2012ApJ...758L...9M,2013ApJ...772..119L, 2015ApJ...800...20G, 2018ApJ...853..179T}, and GC mass fraction from total gas mass (the second tunable parameter). Once the total GC mass at the epochs of formation is fixed, individual GC masses are sampled from a power $-2$ mass function observed from young massive clusters \citep{2010ARA&A..48..431P,2015A&A...582A..93S}. It is encouraging to note that this simple two-parameter model successfully reproduced several key observed properties of GC populations, such as metallicity bimodality, GC mass-halo mass relation, etc. This serves as a much improved model of GC formation for the study of NSCs in galaxies with different assembly histories.

As GCs form at multiple epochs across the galaxy assembly history, their initial spatial distribution varies and influences their subsequent orbital evolution. Thus, we need to carefully take care of this issue at each formation epoch. For GCs formed inside the galaxy ('in-situ'), we assume that, similar to normal star formation, GCs follow similar spatial distribution of cold gas during their formation. As gas does not exhibit a bulge structure, we adopt the continuous spherical S\'ersic density distribution, based on which GCs are assigned to specific galacto-centric radii.
On the other hand, for GCs formed inside satellite galaxies and were brought in via galactic mergers ('ex-situ'), we lack detailed knowledge of the stellar dynamics during galactic mergers. As a simplified treatment, we place these ex-situ GCs at half the virial radius of the descendant halo. This shouldn't affect our GC distribution at the galaxy center at $z=0$, though, because these massive GCs carry large angular momenta relative to the descendant halo, which are highly unlikely to be sufficiently reduced by dynamical friction from the ambient stellar density.

The S\'ersic spatial density distribution was proposed by \citet{1997A&A...321..111P} to match the well-established S\'ersic surface brightness profile \citep{1963BAAA....6...41S}. The spherical density distribution is given by:
\begin{equation}\label{eq:sersic}
\rho(r)=\rho_0 (\frac{r}{R_e})^{-p} e^{-b (r/R_e)^\frac{1}{N_s}},
\end{equation}
Here $\rho_0$ is a normalization, $R_e$ is the effective radius of the galactic disk, $N_s$ is the concentration index. The term $b$ is a function of $N_s$ to ensure half of the projected light is contained within $R_e$, and can be well approximated analytically by $b=2N_s-1/3+0.009876/N_s$ for $0.5<N_s<10$  \citep{1997A&A...321..111P}. The form of $p$ is adopted from \citet{2000A&A...353..873M} as $p=1.0-0.6097/N_s+0.05563/N_s^2$ for $0.6<N_s<10$ and $10^{-2}\leq{r}/R_e\leq10^3$ to match the S\'ersic surface brightness profile \citep{2005MNRAS.362..197T}. 

The effective radius $R_e$ is related to the virial radius of the halo, $R_{\rm vir}$, and halo spin parameter, $\lambda$, by assuming a classical galactic disk formation model (e.g. \cite{1998MNRAS.295..319M}):
\begin{equation}
    R_e = \lambda R_{\rm vir}/\sqrt{2}.
\end{equation}
Since we do not have the information of the total energy of the dark matter halo to estimate the traditional spin parameter defined in \citet{1980lssu.book.....P}, we instead use an alternative definition
$\lambda_{\rm B}=j_{\rm sp}/\sqrt{2}R_{\rm vir}V_{\rm vir}$ that only requires the specific angular momentum and virial velocity $V_{\rm vir}$ (e.g. \cite{2001ApJ...555..240B}). Both $M_{\rm vir}$ and $j_{\rm sp}$ are provided at snapshots of the Illustris simulation, and $v_{\rm vir}$ and $r_{\rm vir}$ are calculated accordingly as explained in the Appendix.

\begin{figure}
	\includegraphics[width=\columnwidth]{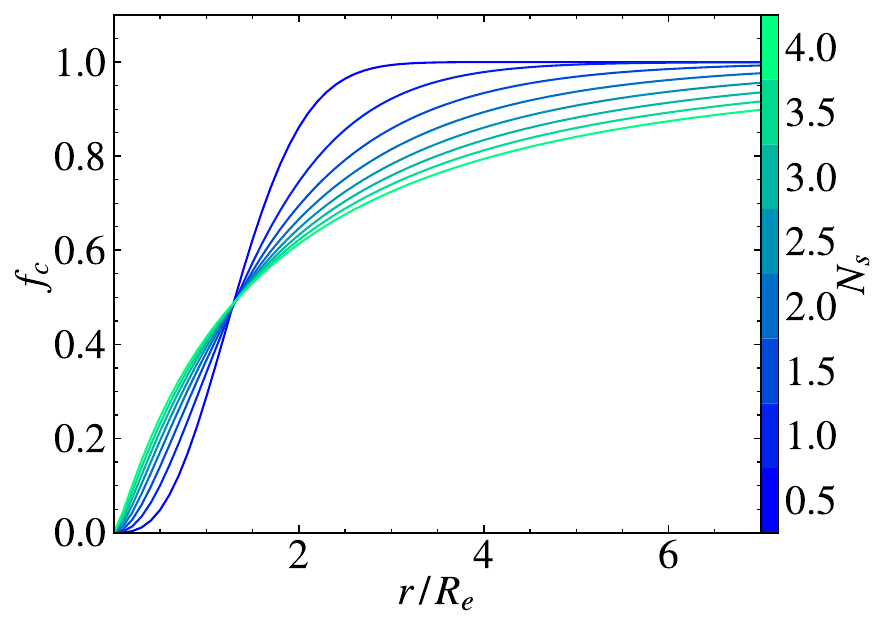}
	\caption{Distribution of the cumulative mass fraction ($f_c$) for the S\'ersic volume density profile (Eqn. \ref{eq:sersic}), plotted against the radial distance $r$ normalized by the effective radius $R_{\rm e}$. Different concentration index $N_{\rm s}$ are plotted in different colors according to the colorbar at the right.}
	\label{fig SersicC}
\end{figure}

Regarding the concentration index $N_s$, larger values were conventionally associated with higher concentration, but we found this to be misleading. In Fig.~\ref{fig SersicC}, we present the cumulative mass fraction distribution versus normalized radial distance for different $N_s$. We see that larger $N_s$ does exhibit higher concentration in the inner region. However, beyond the crossing point at $r\gtrapprox R_e$ ( which we refer to as S\'ersic crossing hereafter), they reach saturation much slower than curves with smaller $N_s$. Consequently, while larger $N_s$ values indicate a more peaked distribution, they also indicate a greater spread. In our subsequent studies, we investigate the influence of different $N_s$ values on our results within the range of 0.5 to 4, with increments of 0.5. As a fiducial model, we select $N_s=2$ following the work of \citet{G+14}.

To study the GC system in the MW and M31, we select Illustris halos with similar masses at $z=0$. For the MW, we adopt the findings by \citet{2021MNRAS.501.5964D}, who estimated a virial mass of 1.01$\pm$0.24$\times 10^{12}M_{\odot}$. Among the Illustris halos, 1099 fall within this range. For M31, there are significant uncertainties, therefore we adopt a wider range of 0.7-2.5 $\times 10^{12}M_{\odot}$ based on the review by \citet{2018MNRAS.475.4043K}. Accordingly, 2029 halos were selected, including those selected for the MW.

\subsection{Dynamical evolution of GCs}
\label{sec gcevo}
After birth, GCs migrate towards the galactic center due to dynamical friction while depositing masses due to stellar evolution and tidal stripping along the way. In situations where they have migrated into the innermost region, the tidal force can be so strong that GCs get completely torn apart \citep{2018MNRAS.477.4423A, 2023ApJ...944..140W}. Thus, GCs that make all the way into the inner a few parsecs can build up NSC and contribute to the MSP populations there. We 
directly adopt the analytical prescriptions of \citet{F+19M31} in modeling: 1) orbital migration, 2) tidal stripping, and 3) direct tidal disruption. These prescriptions include corrections to parameters originally proposed by \citet{G+14} and \citet{F+18MW}. Below, we describe two improvements we made in this work. 

The first improvement is an updated prescription of the mass loss due to stellar evolution, which is obtained from the stellar population synthesis model FSPS \citep{2010ApJ...712..833C}. This allows us to account for the evolving nature of stars within GCs more accurately. 

The second improvement is modeling the time-varying gravitational potential by extracting the time-evolution of the dark matter halos from the Illustris simulation. For any cosmic time, we linearly interpolate the halo mass and spin between adjacent Illustris snapshots, and calculate galactic structural parameters accordingly. The overall gravitational potential comprises three components: the dark halo, the stellar component, and modeled GCs. The dark halo is described using the NFW distribution \citep{NFW}, and the calculations for determining halo parameters are introduced in the Appendix \ref{sec apdx}. The stellar component is modeled with a S\'ersic distribution, as explained earlier. The mass of modeled GCs is also included in calculating the overall potential. For cosmological parameters, we adopt a flat $\Lambda$CDM model with $h=0.704$ and $\Omega_{\rm m,0}=0.2726$, consistent with the Illustris simulation.

To improve the efficiency of our simulation, we implement sub-cycling in the evolution of GCs. We divide the entire time span into 100 sections, each characterized by a constant background potential. To evolve individual GCs, we first calculate their evolution timestep $dt_i$ being the smaller of the tidal and orbital evolution timescales multiplied by a fraction, ${\rm ts_m}$ and ${\rm ts_r}$, respectively. Then, a GC with the smallest value of ${\rm t_i+dt_i}$ means that it evolves the fastest, and exerts the strongest influence on other GCs. Thus, we find and evolve such GC at each step, until all GCs cross the current time span. 

Meanwhile, we found that it is usually the first step of calculated ${\rm dt_i}$ that is overestimating. To efficiently address this, we introduce a maximum cutoff timestep ${\rm dt_{max}}$ on top of the ${\rm ts}$ factors. By testing single GCs with different masses and galacto-centric positions, we optimize the values of ${\rm dt_{max}}$, ${\rm ts_m}$ and ${\rm ts_r}$ together. We find that ${\rm dt_{max}}=0.01\; \rm Gyr$ and ${\rm ts_m=ts_r}=0.2$ strike a balance between efficiency and accuracy.

For calculating the change in GC mass and galactocentric distance in each step, we employ the Runge-Kutta 4th order method. This method offers a significant speed improvement of approximately 20 times compared to the 1st order Euler method.

\subsection{MSPs from GCs} 
MSPs are deposited by GCs due to tidal stripping and disruption, thus they can contribute to the unresolved $\gamma$-ray excess. In calculating the $\gamma$-ray luminosity at $z=0$ from MSPs, we follow \citet{F+18MW,F+19M31} who set the total $\gamma$-ray luminosity of deposited MSPs equal to that of the debris of the GC, using luminosity-mass relation of GCs fitted to observations. To account for uncertainties in this fitting, they also used a constant luminosity-mass relation for comparison, and labeled it as the model 'C'. The model with fitted luminosity-mass relation was labeled 'EQ'.

Then, individual MSP luminosity was sampled according to the observed MSP luminosity function. To account for MSP spin-down due to the loss of rotational energy via magnetic dipole braking, \citet{F+18MW,F+19M31} use two models of the spin-down timescale: a Gaussian distribution model (GAU) and a log-normal distribution model (LON).  In our study, we follow their prescribed methodology and calculate the $\gamma$-ray luminosity at redshift $z=0$ using the four models, namely GAU-C, GAU-EQ, LON-C, and LON-EQ.

\section{RESULTS}
\label{sec results}
In this section, we present our results on the properties of GCs at $z=0$ for MW and M31-like galaxies, and compare with observations. We then show the predicted mass of the NSC and $\gamma$-ray luminosity distribution from MSPs. We also explore how the above quantities vary with different galaxy assembly histories.

\subsection{GC-halo mass scaling relation}
\label{sec mm}
One of the most striking observations of GCs is a linear correlation between the mass of the GC system and its host halo \citep{2009MNRAS.392L...1S, 2010MNRAS.406.1967G,2014ApJ...787L...5H,2017ApJ...836...67H}. The mass ratio is approximately $10^{-5}$ across 5 orders of magnitude \citep{2015ApJ...806...36H}. Therefore, it is crucial for our model to reproduces this scaling relation.

\begin{figure}	 
    \includegraphics[width=0.9\columnwidth,keepaspectratio]{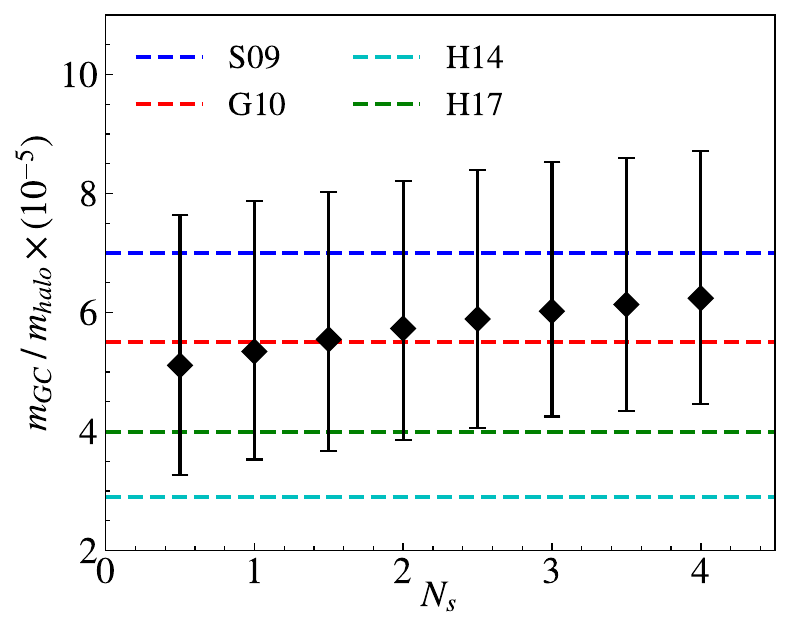}
    \caption{The ratio between the total GC system mass ($m_{\rm GC}$) and the host halo mass ($m_{\rm halo}$) at $z=0$ for different S\'ersic index $N_{\rm s}$ for all selected MW and M31-like halos. The black diamonds show the median ratio, while the error bars show the 25-75 interquartile range. For reference, we also overplot horizontal dashed lines for a compilation of recent observations from \citet{2009MNRAS.392L...1S} (S09, blue), \citet{2010MNRAS.406.1967G} (G10, red), \citet{2014ApJ...787L...5H} (H14, cyan), and \citet{2017ApJ...836...67H} (H17, green).}
	\label{fig mGCmHalo}
\end{figure}

Fig.~\ref{fig mGCmHalo} shows our model prediction of the scaling relation for the MW and M31-like halos combined for different $N_{\rm s}$. We see that the runs with smaller $N_{\rm s}$ tend to produce lower mass in the GC system at $z=0$. This is in line with our findings in Fig.~\ref{fig SersicC}, that a smaller $N_s$ corresponds to a smaller spatial spread of formed GCs, which results in more mass loss due to stronger tidal effect. Nevertheless, the changes of total GC mass in different $N_{\rm s}$ is very small and all choices of $N_{\rm s}$ produce the GC mass in reasonably good agreement with observations. Consequently, our model successfully reproduces the GC-halo mass scaling relation.

\subsection{Spatial distribution of GC number density}
\label{sec nd}
\begin{figure}
	\includegraphics[width=\columnwidth]{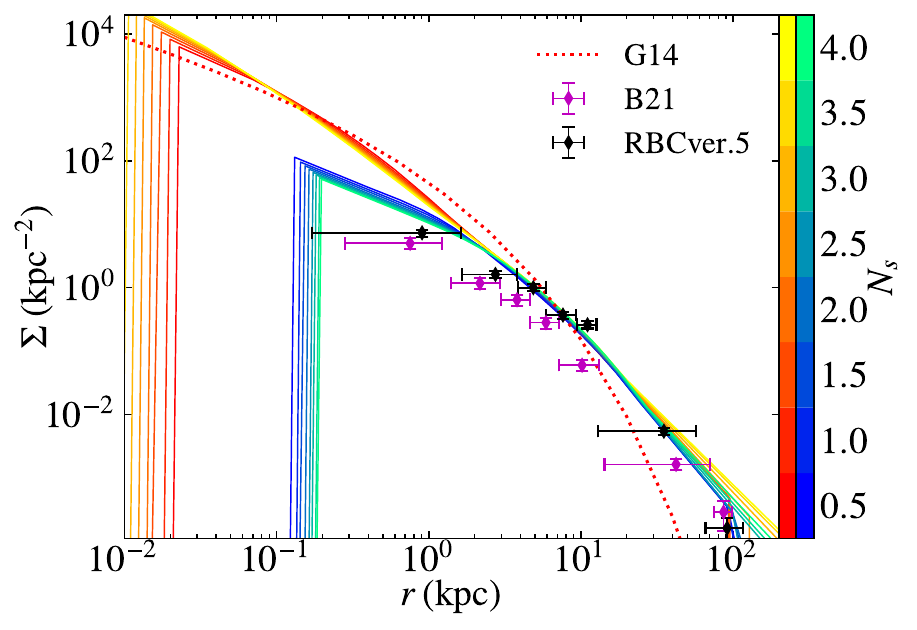}
	\caption{The surface density distribution of GC number with different $N_{\rm s}$ averaged for all selected halos. We do not separate results for MW and M31-like halos, because we have checked that they are barely distinguishable, and that the MW mass range is enclosed by that of the M31. In this plot, the redish curves denote all initial GCs formed at different redshift, and the bluish denote final GCs at $z=0$. We show in red dotted curve (G14) the initial GC distribution in \citet{G+14} for comparison. The error-bars (B21, purple; RBC ver.5, black) show the observed GC catalogue for the MW \citep{B+21} and M31 \citep{2014yCat.5143....0G}. In B21, $\omega Cen$ was manually removed as it was proved to be the stripped core of a disrupted dwarf galaxy \citep{2008ApJ...676.1008N}.}
	\label{fig NDns}
\end{figure}

In this section, we closely examine the spatial distribution of GCs in their number density at $z=0$. Fig.~\ref{fig NDns} shows the average trend for MW and M31 combined, as we have checked that they are barely distinguishable. 

When we compare the initial and final GC distributions, we notice that the number density outside $\approx$3 kpc barely decreases. This can be attributed to the fact that beyond this position, most GCs with a mass of approximately $10^5 M_{\odot}$ have tidal and migration timescales longer than a Hubble time. In addition, stellar evolution alone cannot exhaust a GC. 

When we investigate the effect of different $N_{\rm s}$ values, it turns out that this made minimal difference for both the initial and final GC distributions, except for the innermost region. In the initial distribution, the typical S\'ersic crossing is only observed $\approx$100 pc, since GC formation across cosmic times overlaps in the outer regions as the halo grows, erasing outer crossings. Our inclusion of ex-situ GCs also contributes to this effect. In the final distribution, although the difference between $N_{\rm s}$ values remains minimal, the trend is opposite to that observed in the initial distribution. Larger $N_{\rm s}$ values within the S\'ersic crossing lead to stronger tidal disruption, resulting in fewer surviving GCs and a smaller number of GCs in the final distribution. 

\begin{figure}
	\includegraphics[width=0.9\columnwidth]{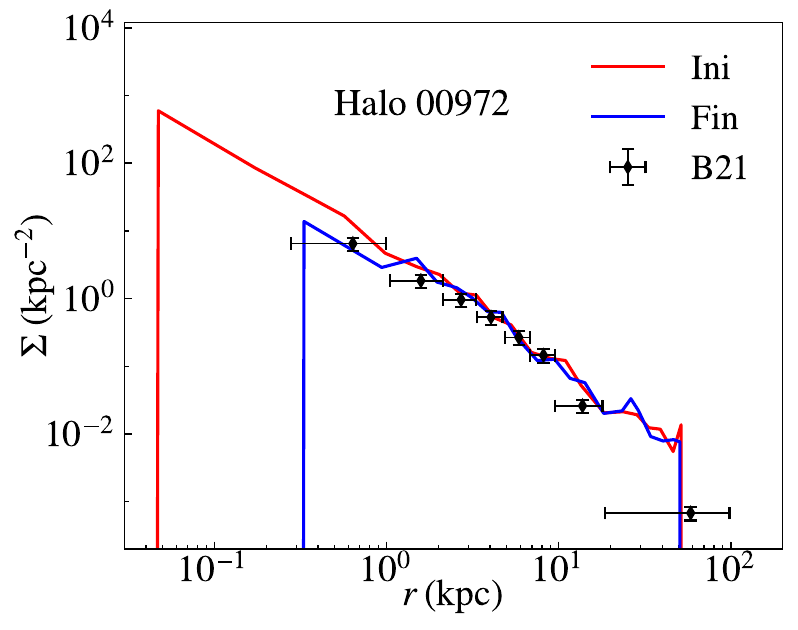}
	\caption{GC number density distribution for one candidate MW halo with Illustris subhalo ID 00972. The red line shows its initial GCs, and blue line for GCs at $z=0$. The error bars are again the \citet{B+21} observed GC catalogue. }
	\label{fig 972nd}
\end{figure}

Comparing our results with observations, we found a minor overshoot within $\approx$ 5 kpc for the MW. However, individual halos exhibit a spread around the average distribution, and several candidate halos demonstrate conformity to the observation. We present one such candidate halo in Fig.~\ref{fig 972nd}. Therefore, our model can be considered an acceptable fit to the observed spatial distribution of MW and M31 GCs. Additionally, we observed no preference for $N_{\rm s}$, so we will keep $N_{\rm s}=2$ as the fiducial choice.

To further analyze the GC population, we utilize the orbital information in the \citet{B+21} catalogue to distinguish between the in-situ and ex-situ sub-populations of MW GCS. In order to do so, we follow \citet{2019A&A...630L...4M}, who analyzed the dynamical differences of the two branches of GCs on the age-metallicity plot for 151 MW GCs. Based on their distinct features, they assigned 62 GCs as having formed in-situ, which comprise of bulge GCs and disk GCs. The former are defined as having apocenter distance less than 3.5 kpc, while the latter as having orbital altitude less than 5 kpc and circularity greater than 0.5. For our GC catalogue, we calculate the orbital and potential parameters using the $Python$ package $Galpot$ by \citet{2017MNRAS.465...76M,1998MNRAS.294..429D}. Since our treatment of ex-situ GCs is rudimentary, we focus on in-situ GCs for now.

\begin{figure}
	\includegraphics[width=\columnwidth]{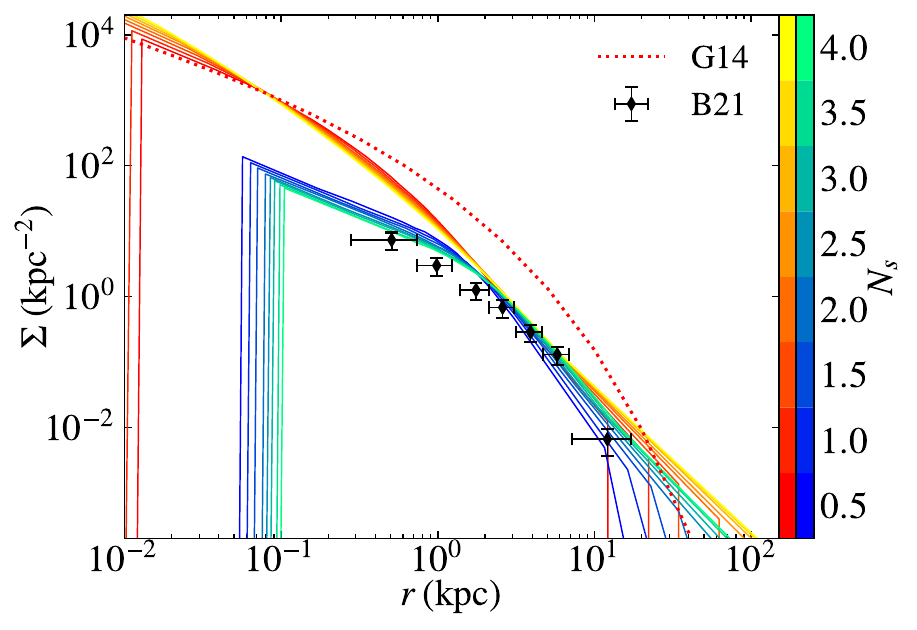}
	\caption{Same as Fig.~\ref{fig NDns}, but only for MW in-situ GCs.}
	\label{fig NDnsI}
\end{figure}

In Fig.~\ref{fig NDnsI}, we plot the number density distributions of MW in-situ GCs in a similar manner to Fig.~\ref{fig NDns}. We can observe that most of the observations made for the overall GC distribution also apply to the in-situ GC distribution, except for 2 differences. First, in-situ GCs exhibit a more concentrated distribution, with a significant drop in their numbers towards the outskirts. This suggests that in-situ GCs preferentially populate the central regions of the MW-like halos. Second, without the contribution from the ex-situ population, the dispersion between different values of $N_s$ becomes more pronounced in the outskirts, and we can observe the outer portion of the S\'ersic crossing. Nevertheless, we can regard our model as satisfactorily reproducing the observed GC number density distribution, both overall and in-situ, and $N_s$=2 as a reasonable choice for the concentration parameter.  

\begin{figure}
	\includegraphics[width=0.9\columnwidth,keepaspectratio]{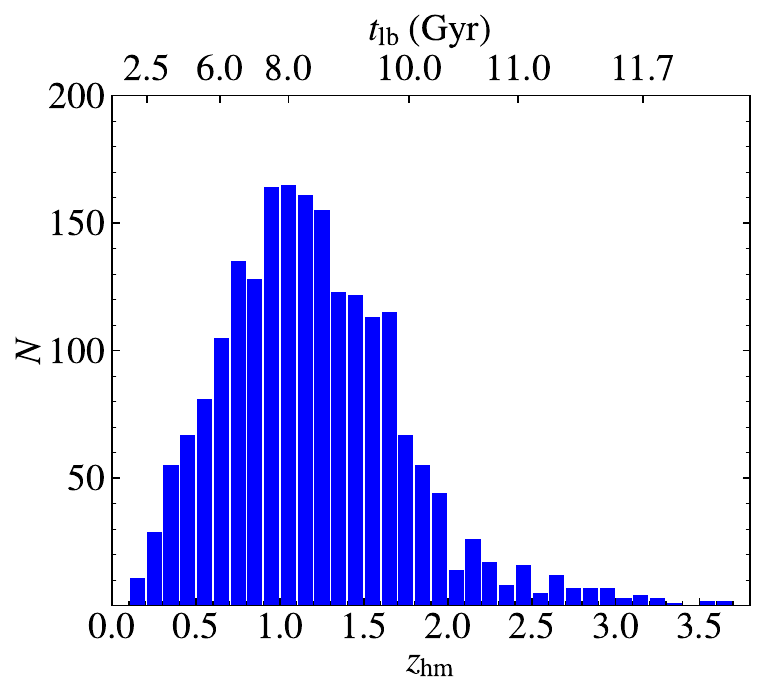}
	\caption{Histogram showing the half-mass redshift $z_{\rm hm}$ for all candidate halos. Look-back time is shown on the upper horizontal axis as well.}
	\label{fig z}
\end{figure}

One advantage of utilizing the halo catalog extracted from Illustris is the ability to investigate the influence of different assembly histories on halos with similar masses. To parameterize the halo assembly histories, we employ the concept of half mass redshift, denoted as $z_{\rm hm}$, which represents the redshift at which the halo acquired half of its present mass. In Fig.~\ref{fig z}, we show the distribution of $z_{\rm hm}$ values. The histogram exhibits a log-normal shape centered around $z\sim1.2$, corresponding to a look-back time of approximately 8 billion years, with a tail extending towards higher redshifts.

\begin{figure}
	\includegraphics[width=\columnwidth]{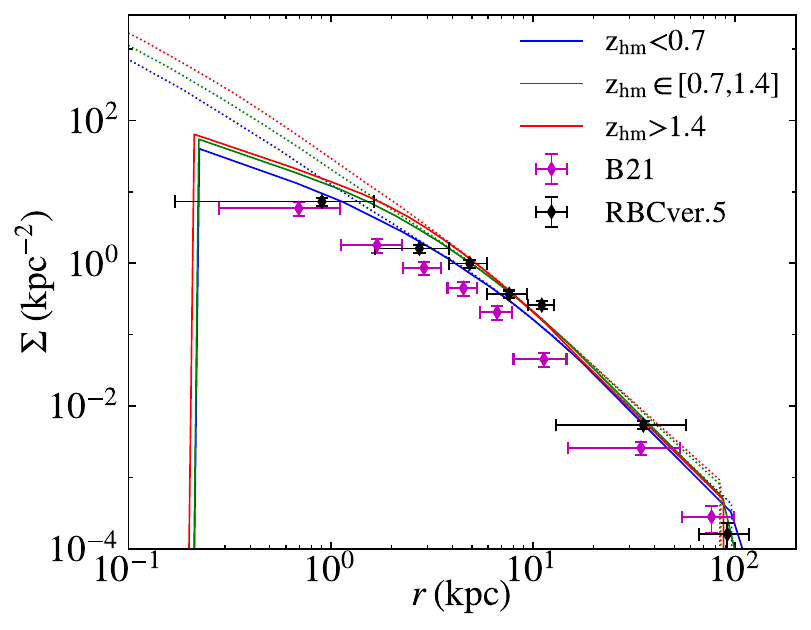}
	\caption{Number density distribution for all candidate halos with different $z_{\rm hm}$. For clarity purposes, we only show the average distribution for three $z_{\rm hm}$ ranges. Dashed lines correspond to initial GCs and solid line to final ones.}
	\label{fig NDz}
\end{figure}

We are interested in whether $z_{\rm hm}$ has any discernible effect on the spatial distribution of GCs. Fig.~\ref{fig NDz} illustrates the GC number density distribution with different $z_{\rm hm}$ for MW and M31-like galaxies. When we look at the influence of $z_{\rm hm}$ for initial GCs, galaxies that formed earlier (referred to as EFGs, earlier formed galaxies) tend to exhibit more concentrated distributions of GCs compared to halos that formed later (referred to as LFGs, later formed galaxies). This trend can be attributed to the higher merger rate of halos at larger redshifts, as indicated by previous studies \citep{2008MNRAS.386..577F, 2010MNRAS.406.2267F}. Consequently, EFGs experience more GC formation events, starting from smaller halo sizes, while LFGs undergo fewer GC formation events, each resulting in significant growth of halo mass and size and leading to a more spread-out distribution of formed GCs.

In the case of final GCs, however, the impact of $z_{\rm hm}$ is less pronounced. This can be understood, as more concentrated GCs also experience stronger tidal disruption and dynamical friction, which reduces their numbers. Additionally, as these GCs formed earlier, these disruptive processes act over a longer period. Consequently, the final GC number density distribution shows minimal traces of the galaxy assembly history.

When comparing our results with observations, as the influence of $z_{\rm hm}$ on the final GC distribution is not significant, the results are similar to those of Fig.~\ref{fig NDns}.

\subsection{GC mass function}
\label{sec mf}
\begin{figure}
    \centering
    \includegraphics[width=0.9\columnwidth]{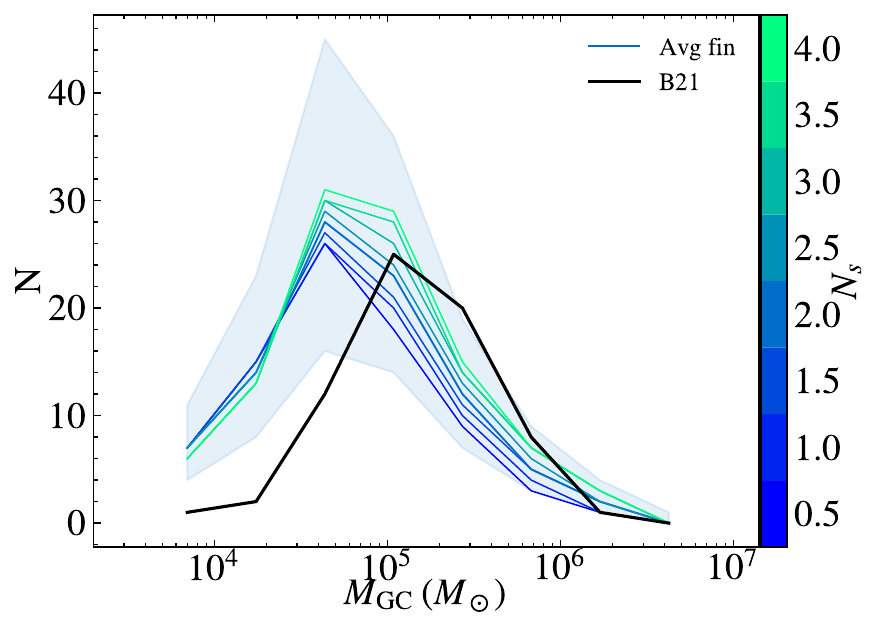}
    \caption{The mass function of in-situ GCs at $z=0$, showing the median of all MW-like halos for different $N_s$. The shaded region corresponds to the 25-75 interquantile range for $N_s$=2. The GC catalogue by \citet{B+21} is plotted in black solid line for comparison.}
    \label{fig masfunmwistns}
\end{figure}

\begin{figure}
    \centering
    \includegraphics[width=0.8\columnwidth]{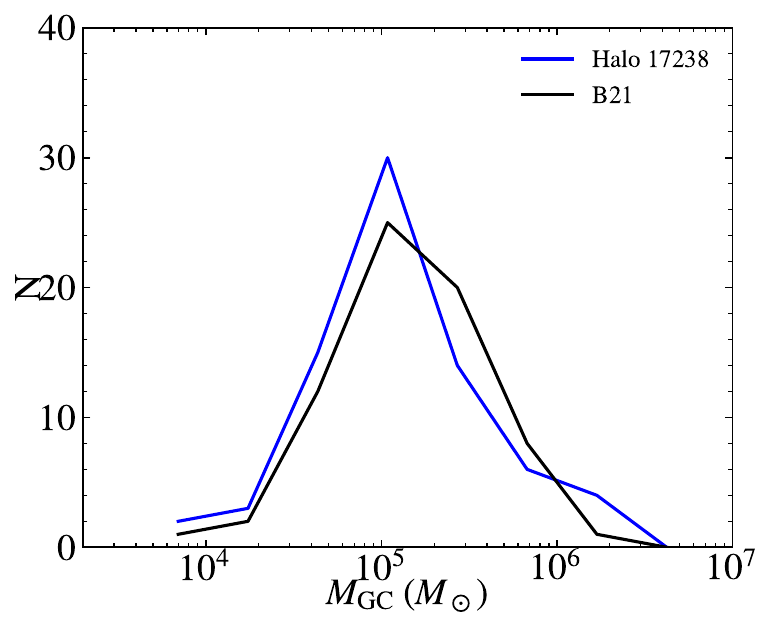}
    \caption{The same as Fig.~\ref{fig masfunmwistns}, but for one candidate halo with Illustris subhalo ID 17238.}
    \label{fig masfunmwistns17238}
\end{figure}

Besides the spatial distribution of GCs, their mass function at $z=0$ is also an important property to compare with observations. Previous studies using the same GC formation model \citep{2010ApJ...718.1266M, LG14, CGL18,CG19,2022MNRAS.514.4736C,2023MNRAS.522.5638C} have carried out comprehensive analyses, confirming a transformation from the initial power-law mass function to a log-normal shape at $z=0$ that agrees with observations. Therefore, in this study we focus on the in-situ GC population. In Fig.~\ref{fig masfunmwistns} we show the median mass function of candidate MW-like halos for different $N_s$, and overplot observation results. We see that our model also shows a log-normal shape, although with its peak shifted towards smaller GC masses compared with observations. This can be due to the incompleteness in observed low mass GCs. Nevertheless, the shaded region corresponding to $N_s=2$ can cover most of the observation trend. Indeed, candidate halos can fit the observation quite well, as illustrated by Fig.~\ref{fig masfunmwistns17238}. Thus, our model provides a good fit to the observed in-situ GC mass function.

We also observe that $N_s$ results in more noticeable discrepancies in the predicted mass function above $\sim 10^5 M_{\odot}$, while at lower masses, the influence of different $N_s$ is flipped and less prominent. Larger $N_s$ leads to slightly fewer light GCs but more heavier ones, and vice versa. This is due to the fact that those light GCs are prone to weak dynamical friction, thus they barely migrate inward. Their disruption is mostly determined by their position at birth. As we have observed in Fig.~\ref{fig SersicC}, larger $N_s$ have peaked density at the galaxy center, which leads to prominent disruption of lighter GCs that initially formed in the region. On the other hand, as the general distribution is more spreaded, heavier GCs have a longer journey to migrate to the galaxy center and are subject to weaker tidal effect. Thus, the distribution leads to slightly stronger disruption of light GCs formed in the innermost region, but weaker disruption of heavier ones.

\subsection{NSC mass} 
\label{sec md}
\begin{figure}
	\includegraphics[width=\columnwidth]{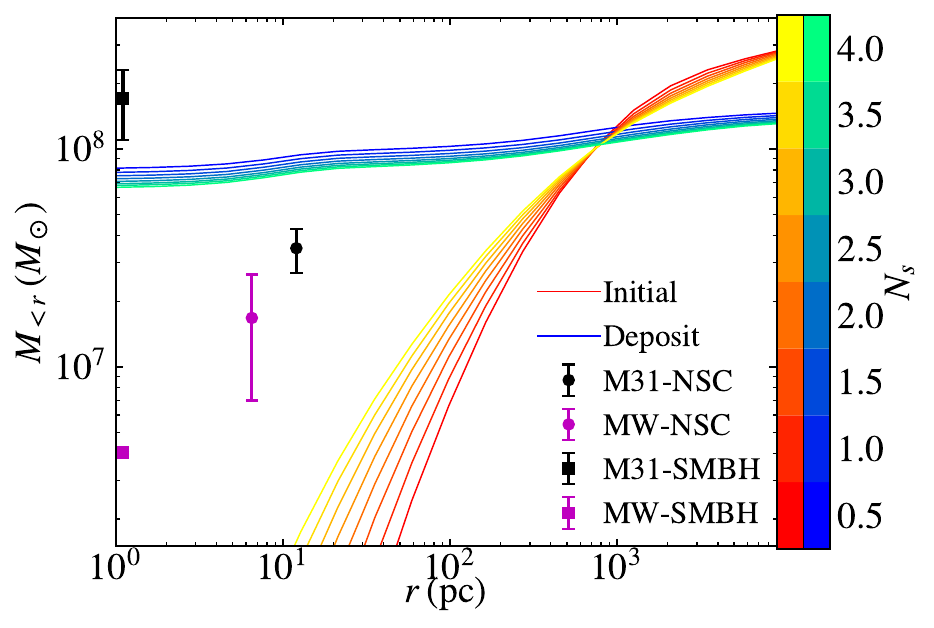}
	\caption{Spatial distribution of the cumulative masses of initial GCs (summed over all formation epochs) and their deposition at $z=0$, averaged for MW and M31-like halos with different $N_{\rm s}$. The color code is identical to Fig.~\ref{fig NDns}. The MW SMBH and NSC observations are adopted from \citet{2019Sci...365..664D,2020A&ARv..28....4N}, and M31 from \citet{2005ApJ...631..280B,2022MNRAS.514.5751L} respectively.}
	\label{fig MRns}
\end{figure}

In this section, We examine the NSC mass contributed by GCs as they spiral in. On average, $\sim80$ and 100 GCs migrated into the NSC of the MW and M31 respectively, with smaller $N_s$ corresponding to a few more GCs. The standard deviation is $\sim30$ and 40 respectively. In Fig.~\ref{fig MRns}, we compare the deposited GC masses with the observed NSC mass at $z=0$. Unlike the number density distribution, the cumulative mass distribution of initial GCs exhibits a prominent S\'ersic crossing in the inner region, because the range of the y axis is much smaller in this case. Regarding the deposited mass, we observe that GCs with smaller values of $N_{\rm s}$ contribute more to the total deposited mass, which aligns with the trend we observed in the GC-halo mass scaling relation discussed in Section \ref{sec mm}. Regardless of the influence of $N_{\rm s}$, the deposited mass is predominantly confined to the central regions. This contrasts with the results presented by \citet{G+14,F+18MW}, where a plateau is established starting from 4 pc outward and continues to rise prominently up to 10 kpc. The disparity arises due to our GC formation occurring throughout the entire assembly history of the halo, resulting in a more concentrated distribution of GCs, as shown in Fig.~\ref{fig NDns}. Consequently, GCs are more susceptible to significant tidal effects and tend to deposit mass towards the galaxy center. 
 
\begin{figure}
	\includegraphics[width=\columnwidth]{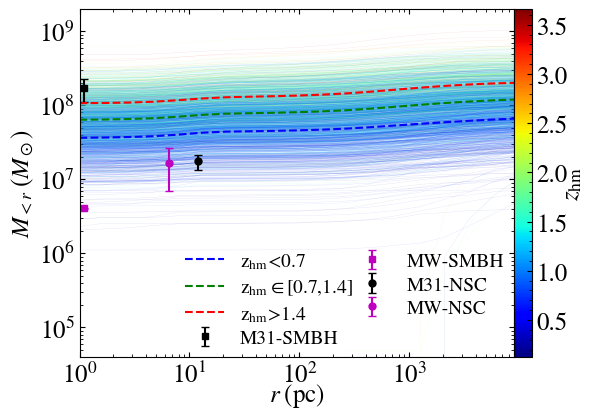}
	\caption{Cumulative spatial distribution of masses of the initial GCs and their deposition with different $z_{\rm hm}$ for all candidate halos. Three dashed curves show the average for halos falling under different $z_{\rm hm}$ spans. Observed positions and masses of the MW and M31 NSC and SMBH are plotted with uncertainties.}
	\label{fig MRz}
\end{figure}

While the average deposited mass exceeds the NSC mass, it is important to note that different galaxy assembly history yield substantial variations, as depicted in Fig.~\ref{fig MRz}. As discussed in the preceding section, EFGs give rise to more GCs, which experience stronger and more prolonged tidal disruption. Consequently, halos with larger $z_{\rm hm}$ have GCs depositing more mass towards the center. Conversely, LFGs exhibit less deposited mass, implying that our MW NSC plausibly originates from a halo with $z_{\rm hm}\sim1$.

\subsection{Gamma ray luminosity}
\label{sec lum}
\begin{figure}
	\includegraphics[scale=0.6]{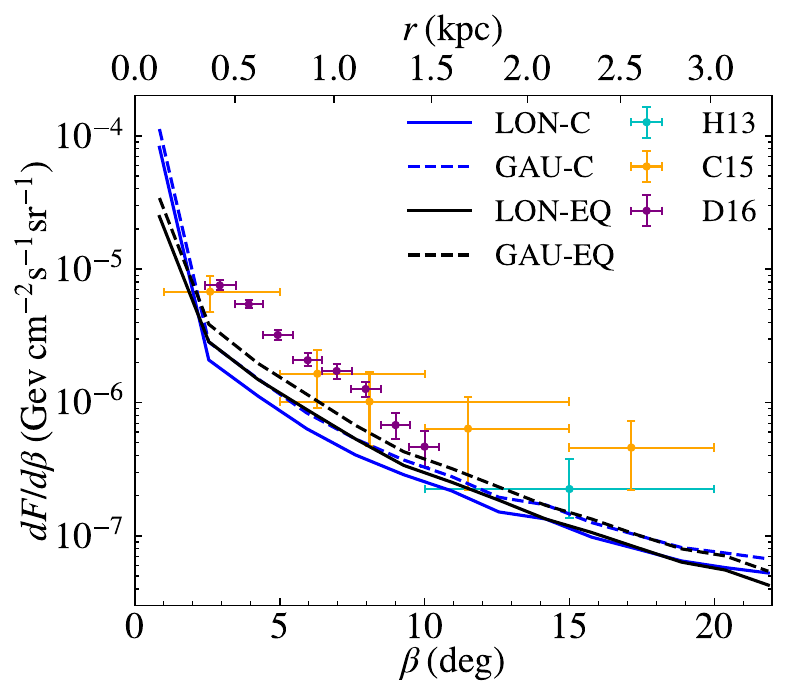}
	\caption{Differential flux distribution over angular distance from the MW center ($\beta$) for the four MSP luminosity models. We overplotted observational constraints by \citet{H+13} (H13, cyan), \citet{C+15} (C15, gold) and \citet{D+16} (D16, purple).}
	\label{fig dL}
\end{figure}

\begin{figure}
	\includegraphics[scale=0.6]{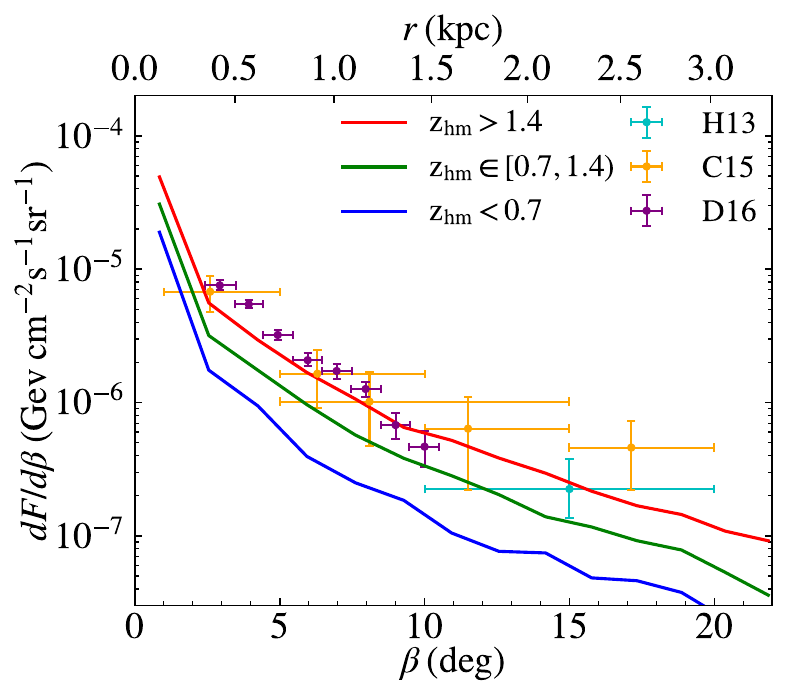}
	\caption{Similar to Fig.~\ref{fig dL}, but for the GAU-EQ model with three $z_{\rm hm}$ ranges.}
	\label{fig dLz3}
\end{figure}

As our GC fittings turn out consistent with observations, we move forward with $N_{\rm s}=2$ to check the spatial distribution of the luminosity of deposited MSPs. As different authors analyzed the $\gamma$-ray excess data in different methods, we present our results in both differential and cumulative distribution. Fig.~\ref{fig dL} shows our model prediction of the differential flux distribution. Notably, different models yield similar results. And although they generally underestimate the excess flux, the overall shape is consistent. Consequently, we select the GAU-EQ model as the best choice and plot it for different $z_{\rm hm}$ intervals in Fig.\ref{fig dLz3}. We observe that EFGs exhibit relatively higher flux emission, as their GCs deposit more MSPs. Halos with $z_{\rm hm}\gtrsim0.7$ provide a good fit to the observations. Note that for clarity purpose, the color code here does not precisely match the colorbar in Fig.~\ref{fig MRz}. With that in mind, we notice that the results are compatible, indicating that a halo with $z_{\rm hm}\gtrsim0.7$ can successfully reproduce both the NSC mass and the spatial distribution of the $\gamma$-ray differential flux emmision.

\begin{figure}
	\includegraphics[scale=0.6]{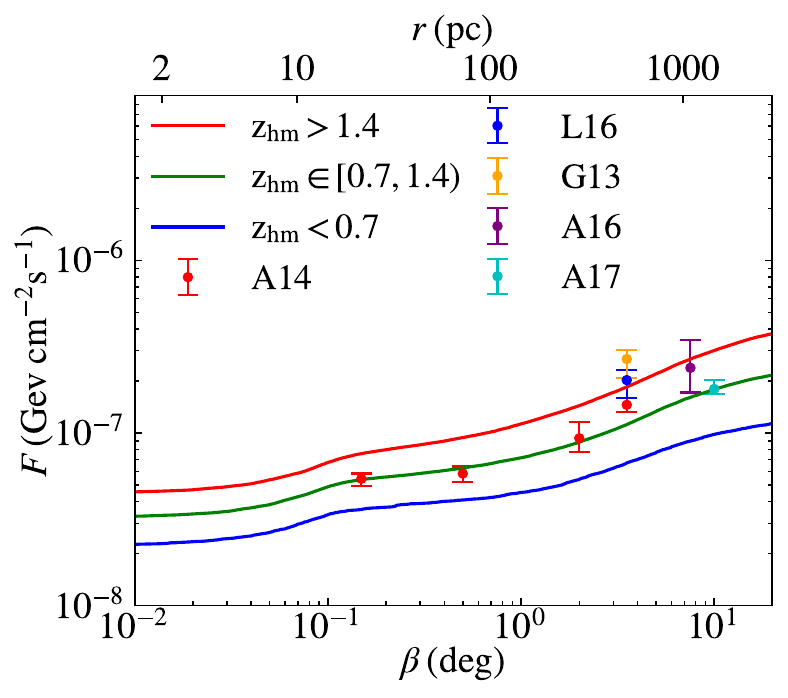}
	\caption{Cumulative flux distribution for the GAU-EQ model with three $z_{\rm hm}$ ranges. Observational constraints are from \citet{A+14} (A14, red), \citet{L+16} (L16, blue), \citet{G+13} (G13, gold), \citet{A+16} (A16, purple), \citet{A+17} (A17, cyan).}
	\label{fig cLz3}
\end{figure}

For the cumulative flux distribution, Fig.~\ref{fig cLz3} illustrates the GAU-EQ model for different $z_{\rm hm}$. We can observe that our model also shows a consistent shape with the observations, which appears better than the results by \citet{F+18MW} as we exhibit more flux in the innermost region. This arise from our GC formation occurring throughout the entire assembly history of the halo. As a result, GCs started their evolution closer to the galaxy center. Once again, halos with $z_{\rm hm}\gtrsim0.7$ provide a good fit. And consequently, it is plausible that the GCE arose solely from deposited MSPs.

\begin{figure}
	\includegraphics[scale=0.6]{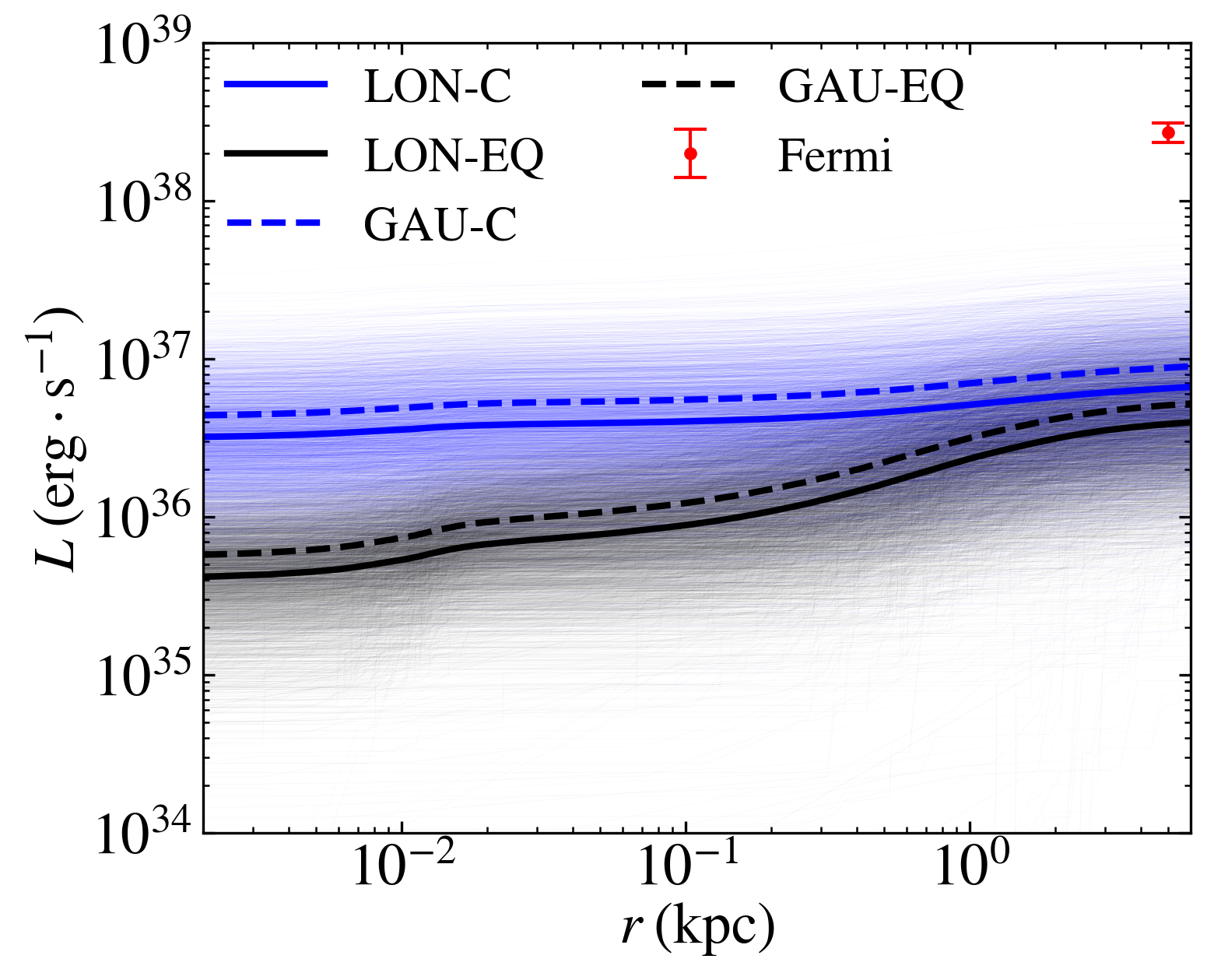}
	\caption{Cumulative luminosity distribution for the four MSP luminosity models for all 2029 M31-like halos. The average distributions for each model are also plotted. The Fermi data was adopted from \citet{F+19M31} in red errorbar.} 
	\label{fig cL31}
\end{figure}

Having examined the GCE, we proceed to the M31 excess. Due to the lack of detailed analysis regarding the excess spatial distribution, we solely compare the cumulative luminosity at 6 kpc. Fig.~\ref{fig cL31} illustrates the cumulative luminosity distribution for the four MSP luminosity models. The models exhibit notable discrepancies, particularly in the innermost region, which diminish as we move towards the outskirts and nearly vanish at 6 kpc. This discrepancy stems from the fact that the EQ models employ a fitted relation, where log$(L_{\gamma}/m_{GC})$ is negatively proportional to log$(m_{GC})$. Consequently, heavier GCs are dimmer compared to the C models that utilize a constant luminosity-mass ratio. Since heavier GCs are more susceptible to dynamical friction and tidal disruption, they primarily contribute to the $\gamma$-ray emission in the innermost regions. As we move farther away from the galaxy center, lighter GCs become more dominant, which reduces the discrepancy. Nevertheless, all models converge and fall short of matching the excess signal at 6 kpc. Additionally, the highest luminosity among individual halos only reaches approximately $7\times10^{37}$ erg/s, which is less than one-third of the excess luminosity. Therefore, it is unlikely that MSPs alone account for the M31 excess. This suggests that the two galaxies have distinct origins of the excess emissions despite their similar masses.

\section{Discussion}
\label{sec discus}
\subsection{Galaxy assembly history}
\label{zhm}

We have observed in preceding sections that galaxy assembly history has clear influences on the GC and NSC properties. We shall take a systematic look in this section. 

\begin{figure}
	\includegraphics[width=\columnwidth]{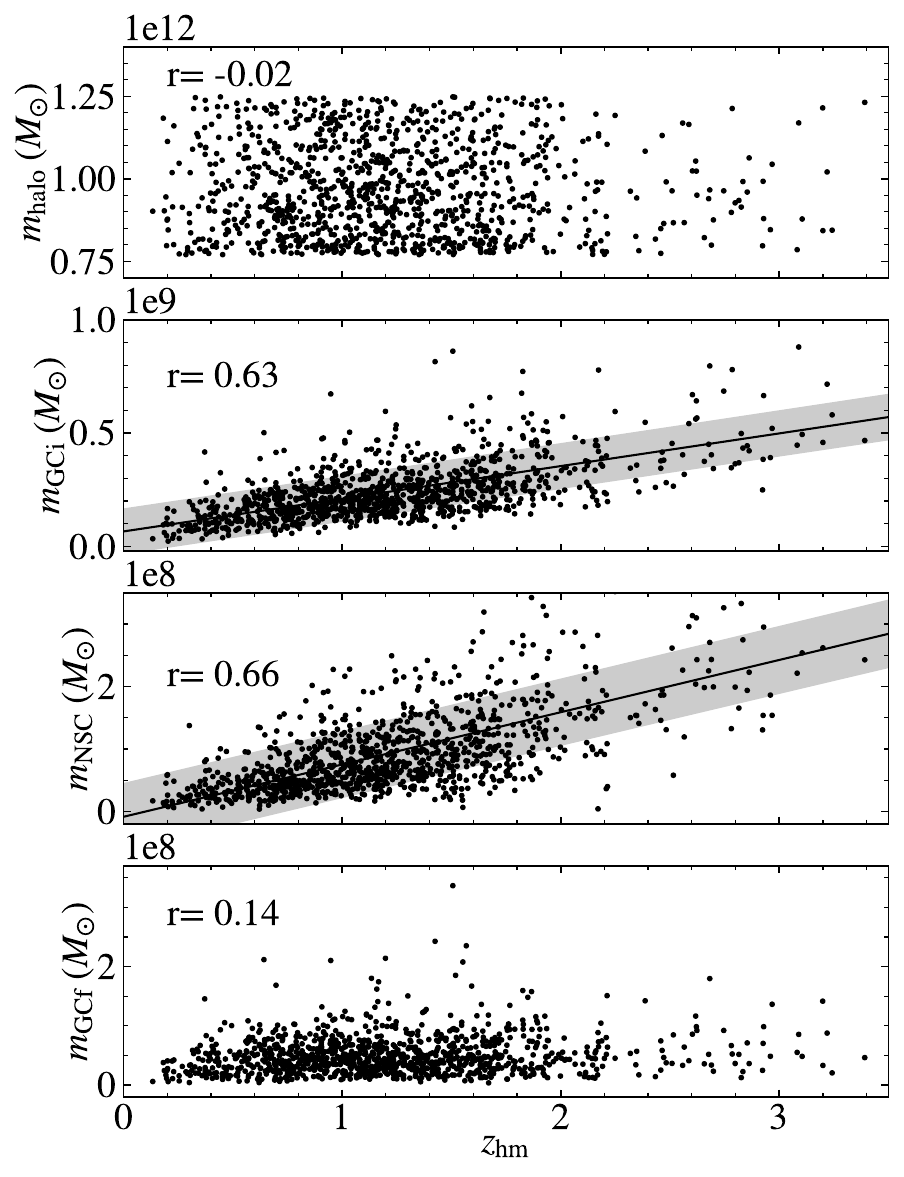}
	\caption{Correlations of $z_{\rm hm}$ with (from the top panel to the bottom) the mass of the halo, initial GCs, NSC and final GCs for all MW-like halos. Pearson correlation coefficients are shown at the upper left corner of each panel. Solid lines and shaded regions show the best-fit trend and 1-$\sigma$ dispersion. The standard deviations of the four masses are 1.38$\times10^{11}$, 1.30$\times10^8$, 4.82$\times10^7$ and 3.33$\times10^7 M_{\odot}$ respectively, which corresponds to 0.14, 0.54, 0.76 and 0.66 of the respective mean.} 
	\label{fig XxZhm}
\end{figure}

In Fig.~\ref{fig XxZhm} we illustrate the correlations between $z_{\rm hm}$ and four important masses: the halo mass, NSC mass, total initial GC mass and GC mass at $z=0$. The halo mass does not correlate with $z_{\rm hm}$ since halos of any mass can assemble early or late. However, the density of data points varies across $z_{\rm hm}$, consistent with the log-normal distribution presented in Fig.~\ref{fig z}. And we have demonstrated previously that EFGs give rise to more GCs, resulting in a positive linear trend between the initial GC masses and $z_{\rm hm}$. We have also showed that EFGs have GCs distributed closer to the galaxy center, making them susceptible to stronger and prolonged tidal disruption. Consequently, they contribute a greater amount of mass to the NSC, leading to a positive linear trend between the NSC mass and $z_{\rm hm}$. Thus, the NSC mass serves as a good indicator in breaking the degeneracy to infer the galaxy assembly history. Larger NSC masses indicate relatively earlier accumulation of the halo mass, and vice versa. However, it is intriguing that $z_{\rm hm}$ appears to provide little information about the final GC mass. This seems to suggest a lack of correlation between initial and final GC masses either, contrary to what one might expect.

\begin{figure}
	\includegraphics[width=\columnwidth]{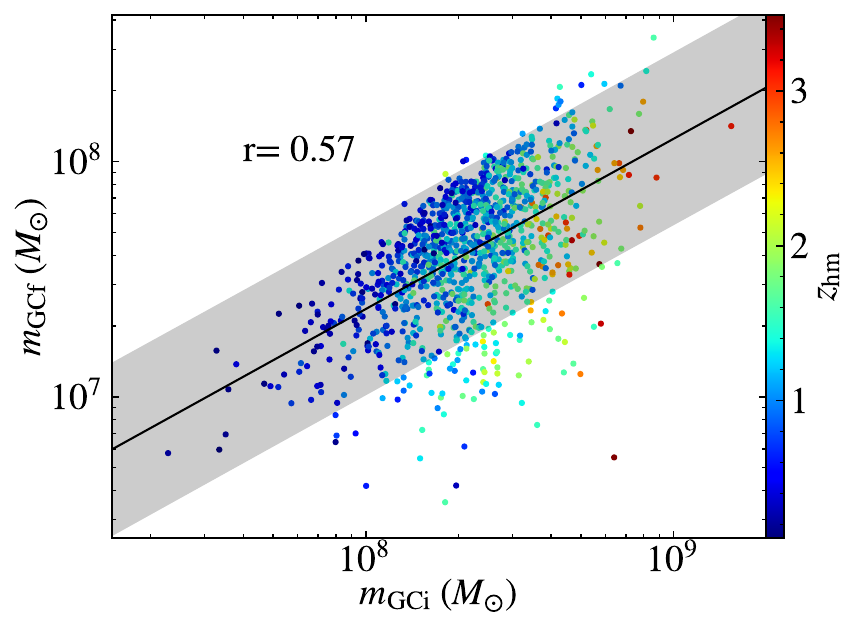}
	\caption{GC initial and final masses for all MW-like halos, with colors denoting $z_{\rm hm}$. The best-fit trend with 1-$\sigma$ band is plotted in black solid line and shaded region.}
	\label{fig GCfiZhm}
\end{figure}

To verify this observation, we show in Fig.~\ref{fig GCfiZhm} the initial and final GC masses for all MW-like halos, with colors denoting $z_{\rm hm}$. A positive trend reasonably exist between initial and final GC masses, although there is relatively large dispersion, particularly at smaller final GC masses. If we observe $z_{\rm hm}$ across different initial GC masses, larger GC masses do correspond to larger $z_{\rm hm}$. However, if we examine final GC masses, each value is associated with a large spread of $z_{\rm hm}$ values. This is due to the fact that the earlier-formed larger GC masses also suffer from stronger disruption, potentially resulting in smaller final masses at $z=0$. Consequently, small final GC masses can arise from either small initial masses or earlier-formed large initial masses. This breaks the correlation between final GC masses and galaxy assembly history. 

\begin{figure}
	\includegraphics[width=\columnwidth]{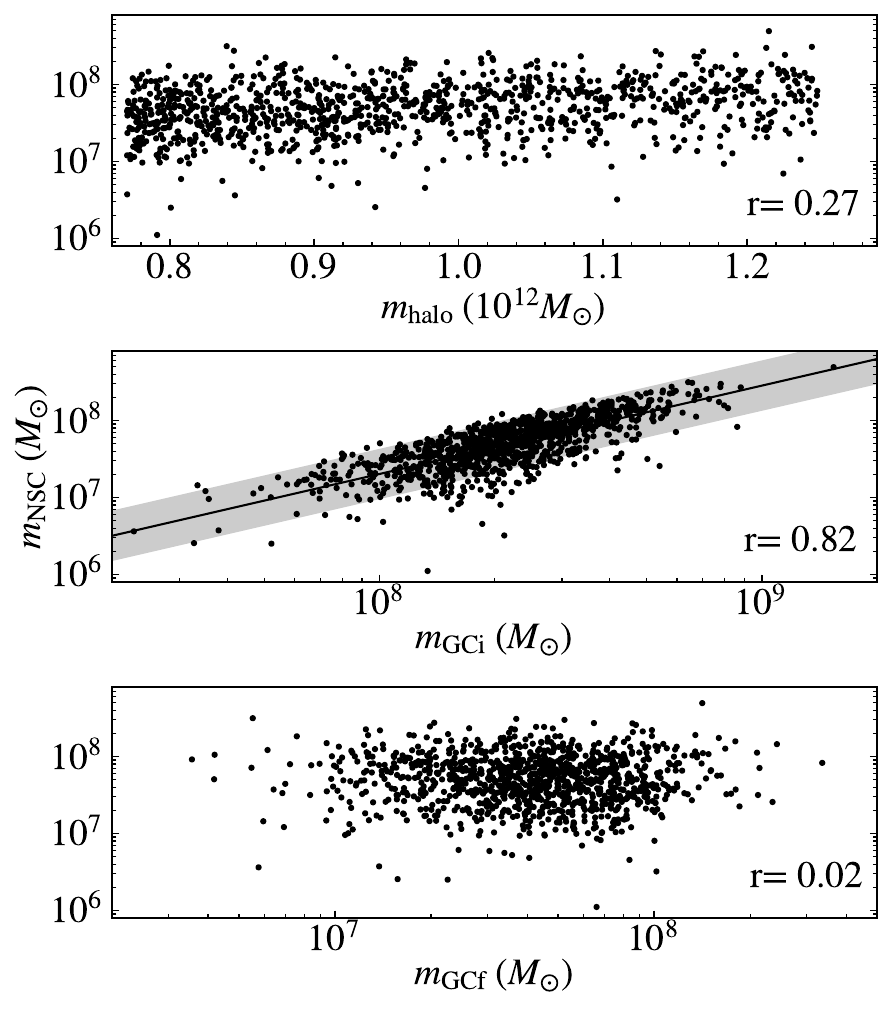}
	\caption{Correlations of the NSC mass with the halo mass (upper panel), total mass of initial GCs (middle panel) and that of final GCs (lower panel) for all MW-like halos. Pearson correlation coefficients are shown at the lower right corner of each panel, and best-fit trend and 1-$\sigma$ range was shown as well. Only the initial GC mass exhibits a positive trend with the NSC mass.}
	\label{fig NscXx}
\end{figure}

We are also interested in correlations associated with the NSC mass, a highly significant outcome of our GC model. Fig.~\ref{fig NscXx} shows the NSC mass plotted against the halo mass, initial GC mass, and final GC mass. As previously demonstrated in Fig.~\ref{fig XxZhm}, the NSC mass is positively correlated with $z_{\rm hm}$, which has no relation to the halo mass. It is not surprising to observe that the NSC mass does not correlate with the halo mass. It is the earlier assembly of the halo that give rise to a heavier NSC, rather than the mass of the halo itself. However, it is worth noting that this observation may change when examining a broader range of halo masses, which will be investigated in our future studies.

In the middle panel of Fig.~\ref{fig NscXx}, a strong positive trend is observed between the mass of initial GCs and the NSC mass. Not only do GCs serve as the fuel for the build-up of the NSC, but larger initial GC masses also correlate with larger $z_{\rm hm}$ values, indicating earlier formation and more concentrated distribution. Collectively, these factors contribute to a larger NSC mass. 

However, final GC masses exhibit no correlation with the NSC mass. This can be comprehended, as we have demonstrated in Fig.~\ref{fig XxZhm} that final GC masses do not correlate with $z_{\rm hm}$. Each GC mass at $z=0$ might arise from either early-formed large $GC_i$ which builds up a large NSC, or later-formed lighter $GC_i$ which contributes little to the NSC mass. Thus, the final GC mass does not serve as an indicator for the NSC mass.

In summary, we presented in this section various correlations associated with the galaxy assembly history. We found that $z_{\rm hm}$, the total initial GC mass and the NSC mass are correlated, due to the fact that EFGs give rise to an old, heavy and concentrated GC system which contributes a larger amount of mass to the NSC. Thus, either value among $z_{\rm hm}$, the initial GC mass and the NSC mass serves as an indicator of the other two. And one important observation is that we could utilize the NSC mass to infer knowledge on galaxy assembly history. Intriguingly, the final GC mass is not correlated with $z_{\rm hm}$ or the NSC mass. This stems from the degeneracy in the relation between initial and final GC masses. 

\subsection{Possible explanations of the M31 gamma-ray excess}
While our results fall short of the M31 excess by an order of magnitude on average, candidate halo can reach approximately one third of the signal at its highest. Nevertheless, even with in-situ MSPs combined, the MSP channel alone is unable to fully explain the M31 excess. Thus, it is evident that a contribution from DM is necessary. 

Upon the first report on the detection of the M31 excess, \citet{2017ApJ...836..208A} have brought up the possible explanation of DM. A primitive estimate inferred from a DM-origin GCE results in a flux deficit by 5 times, though the level of uncertainty was high. 
Subsequent investigations claimed to match the excess luminosity, but commonly identified tensions with observational constraints, such as the under-detection of DM emission in MW dwarf galaxies \citep{2019PhRvD..99l3027D} and a lack of DM radio emission for M31 \citep{2018PhRvD..97j3021M}. Additionally, \citet{2018PhRvD..97j3021M} found that the two preferred DM annihilation channels for M31, namely $b\bar{b}$ and an even mixture of $b\bar{b}/\tau^+\tau^-$, favor smaller DM masses compared to those suggested by the GCE. Consequently, it was suggested that DM alone does not explain the M31 excess. Combining these findings with our results, it becomes clear that a combination of MSPs and DM offers a promising and potentially inevitable way for explaining the M31 excess. However, the question remains as to why the MW and M31 have different origins for producing such excess emission.  

\subsection{Caveats and future works}
In this section discuss the caveats in our model and improvements to be made in future works.

First, although we adopt the analytical expression of tidal disruption from \citet{F+19M31}, it was based on a static spherical galactic background with circular cluster orbits \citep{2008MNRAS.389L..28G}. Our inclusion of the assembly history of galaxies, however, indicates a more complicated galactic background, especially at large redshifts when galactic mergers were more frequent. This complication is twofold. On one hand, the galactic background keeps varying, although in out treatment of linear interpolation, the variation is steady. Thus, the static approximation is not unreasonable. On the other hand, the process of galactic mergers, especially major mergers, perturbs the galactic environment and GC orbits. However, due to the vibrant nature and lack of knowledge on the process, we leave it to future research. The eccentricity of GC orbits also differs from the circularity approximation. However, to accurately capture realistic GC orbital evolution along galaxy assembly histories can be computationally expensive (e.g. \cite{2017ApJ...834...69L,2018ApJ...861..107L,2022MNRAS.514.4736C}. We could only treat our method as a time averaged approximation to real eccentric GC orbits. However, as very eccentric orbits usually happens for ex-situ GCs, the approximation does not significantly affect our results on the properties of the NSC. With improving powers of N-body simulation on galactic mergers and dynamical friction, more knowledge will enable us to incorporate these effects into a more comprehensive model.

In investigating the formation of NSCs, we didn't take into account the in-situ channel of young stars forming in nuclear regions, despite various observational evidence as mentioned in \ref{sec intro}. Our model partially takes care of this channel, as the GC spatial distribution at formation can sometimes sample GCs at the center of the host galaxies. Nevertheless, a more systematic investigation is warranted to obtain a thorough picture of NSC formation.

Besides fitting the NSC mass, future works will be carried out on investigating other NSC properties, such as the age/metallicity distribution and the internal mass profile of the NSC. Besides MW and M31-like galaxies, we will extend our investigation to a broader galaxy mass range and to different galaxy types.

There are also caveats in modeling the MSPs, as their distribution and evolution in GCs remains highly unknown to us. As stellar density increases towards the GC center, LMXBs and MSPs are supposed to peak near the center. There are evidences both from observations and simulation (see \citet{2019ApJ...877..122Y} and references therein). Thus, the number of MSPs stripped and deposited to the ambient environment depends on the current size of the GC. Our treatment essentially assumes a uniform distribution of MSPs in GCs, which only serves as a lower limit to the MSP contribution to the galaxy center $\gamma$-ray excess. On the other hand, following \citet{2019ApJ...877..122Y}, \citet{2022ApJ...940..162Y} studied the $\gamma$-ray excess problem by depositing MSPs only when the GC is fully disrupted. This methodology goes to another extreme by assuming all MSPs residing at the very center. Further knowledge on the distribution of MSPs inside GCs will enable us to better evaluate their $\gamma$-ray contribution to the galaxy center.

Furthermore, we follow \citet{F+18MW,F+19M31} to assume the effect of new MSP formation canceling MSP spindown in GCs, which is somewhat arbitrary. More knowledge is required to properly evaluate these competing factors.

\section{CONCLUSIONS}
\label{sec concl}
In this study, we have presented a comprehensive model of GC formation and evolution, based on the premise that they primarily form following periods of rapid halo mass accretion. Leveraging the results from the Illustris cosmological simulation, we sample GCs across the galaxy assembly history and simulate their subsequent evolution, accounting for the mass loss and radial migration within an evolving galactic background. Our model successfully reproduces key observations of the MW and M31 GC system at $z=0$, including the mass scaling of the total GC system with the host halo, and the spatial distribution of GC number density. For the MW, we also reproduced the spatial distribution for the in-situ subpopulation. 

With this model at hand, we investigate the spatial distribution of deposited masses of migrated GCs to study its link to the formation of the NSC and galaxy center $\gamma$-ray excess. We find that both NSC masses of the MW and M31 can be reproduced. Detailed spatial distribution of the GCE can also come entirely from deposited MSPs. However, the M31 excess strength is three times as large as our most luminous candidate galaxy. Even factoring in in-situ MSPs born at the galaxy center, the MSP channel still cannot fully account for the excess emission. It becomes evident that DM must play a role in explaining the M31 excess, highlighting a fundamental astrophysical difference between the two galaxies. This constitutes another big difference between them, apart from their galaxy center SMBHs differing in mass by about 50 times. Further investigations are demanded in figuring out the causes to and possible links between these differences.

Another intriguing aspect we discovered is the influence of galaxy assembly history on galaxy properties, which we investigated using halo half mass redshift $z_{\rm hm}$. Interestingly, we found that it does not correlate with halo mass, but conveys valuable information about the GC system and NSC mass. Specifically, EFGs with large $z_{\rm hm}$ give rise to an old, heavy and concentrated GC population as they formed, and vice versa. This results in more deposited mass from GCs and a heavier NSC, which in turn serves as an informative indicator of galaxy assembly history.

In conclusion, our comprehensive model of GC formation and evolution provides a robust framework for understanding the properties of the GC system, the NSC and galaxy center $\gamma$-ray emissions. Additionally, our study unveils the significance of galaxy assembly history in shaping galaxy properties. However, the need to invoke DM to explain the M31 excess emphasizes the distinct astrophysical origins of these high energy emissions from the two similar galaxies. Further investigations are warranted to unravel the precise mechanisms that drive these differences and establish a comprehensive understanding of galaxy formation and evolution.

\section*{Acknowledgement}
\label{sec thx}
Yuan Gao would like to thank the University of Hong Kong for providing postgraduate scholarship and essential remote access to online academic resources for conducting this study during the pandemic.   

\section*{Data Availability}
The data underlying this article will be shared on reasonable request to the corresponding author. The datasets were derived from the Illustris simulation results in the public domain: [The Illustris Collaboration, \url{https://www.illustris-project.org/data/}]. The observation data of the MW and M31 are available via corresponding references in the article.

\bibliographystyle{mnras}
\bibliography{main}

\appendix
\section{Determining halo parameters}
\label{sec apdx}
The NFW profile is described in \citet{NFW} as:
\begin{equation}
    \rho_{\rm NFW}(r)=\rho_0 (\frac{r}{R_s})^{-1}(1+\frac{r}{R_s})^{-2} 
\end{equation}
The corresponding potential is given by:
\begin{equation}
    \Phi_{\rm NFW}(r)=-\frac{4 \pi \rho_0 R_s^3 G}{r} ln(1+\frac{r}{R_s})
\end{equation}
Here $\rho_0$ is normalized by the virial mass $M_{\rm vir}=4 \pi \rho_0 R_s^3[ln(1+c)-\frac{c}{1+c}]$, where the halo concentration $c=R_{\rm vir}/R_s$ is taken from \citet{2008MNRAS.391.1940M} with the analytic form: 
\begin{equation}
    c=9.354\;\left(\frac{M_{\rm vir}\,h}{10^{12}M_{\odot}}\right)^{-0.094}
\end{equation}
The virial radius $R_{\rm vir}$ is mapped from $M_{\rm vir}$ and $z$ via:
\begin{equation}
    R_{\rm vir}=\frac{163}{(1+z)\,h} \left(\frac{M_{\rm vir}h}{10^{12}M_{\odot}}\right)^{\frac{1}{3}}(\frac{\Delta_{\rm vir}}{200})^{-\frac{1}{3}} \Omega_{m,0}^{\frac{1}{3}}\;\;\rm kpc
\end{equation}
here $\Delta_{\rm vir}$ is the average halo over-density at $R_{\rm vir}$, which we take from the spherical collapse model as: 
\begin{equation}
    \Delta_{\rm vir}=\frac{18\pi^2+82x-39x^2}{x+1},           \;x\equiv\Omega_m(z)-1
\end{equation}
In our adopted cosmological model, the evolution of matter density parameter equals \citep{2010gfe..book.....M}:
\begin{equation}
    \Omega_m(z)=\frac{\Omega_{m,0}(1+z)^3}{\Omega_{\Lambda,0}+\Omega_{m,0}(1+z)^3}
\end{equation}

\bsp	% typesetting comment
\label{lastpage}
\end{document}